\documentclass[aps,twocolumn,showpacs,preprintnumbers,amsmath,amssymb,superscriptaddress,floatfix,nofootinbib]{revtex4}

\usepackage{mathrsfs,bm}
\usepackage{longtable,lscape}
\usepackage{txfonts}
\usepackage{amssymb}
\usepackage{indentfirst}
\usepackage{graphicx,,booktabs}
\usepackage{multirow}
\usepackage{color}
\usepackage{amssymb}
\usepackage{subfigure}

\usepackage{footnote}

\begin{document}

\title{Faddeev fixed center approximation to $\pi \bar{K} K^*$ system and the $\pi_1(1600)$}

\author{Xu Zhang}
\affiliation{Institute of modern physics, Chinese Academy of
Sciences, Lanzhou 730000, China} \affiliation{University of Chinese
Academy of Sciences, Beijing 101408, China}

\author{Ju-Jun Xie~\footnote{Corresponding author}} \email{xiejujun@impcas.ac.cn}
\affiliation{Institute of modern physics, Chinese Academy of
Sciences, Lanzhou 730000, China} \affiliation{University of Chinese
Academy of Sciences, Beijing 101408, China} \affiliation{Research
Center for Hadron and CSR Physics, Lanzhou University and Institute
of Modern Physics of CAS, Lanzhou 730000,China}

\author{Xurong Chen}
 \affiliation{Institute of modern physics, Chinese
Academy of Sciences, Lanzhou 730000, China} \affiliation{University
of Chinese Academy of Sciences, Beijing 101408, China}
\affiliation{Research Center for Hadron and CSR Physics, Lanzhou
University and Institute of Modern Physics of CAS, Lanzhou
730000,China}

\date{\today}

\begin{abstract}

We investigate the three-body system of $\pi \bar{K} K^*$ by using
the fixed-center approximation to the Faddeev equation, taking the
interaction between $\pi$ and $\bar{K}$, $\pi$ and $K^*$, and
$\bar{K}$ and $K^*$ from the chiral unitary approach. The study is
made assuming scattering of a $\pi$ on a $\bar{K} K^*$ cluster,
which is known to generate the $f_1(1285)$ state. The resonant
structure around $1650$ MeV shows up in the modulus squared of the
$\pi$-$(\bar{K} K^*)_{f_1(1285)}$ scattering amplitude and suggests
that a $\pi$-$(\bar{K} K^*)_{f_1(1285)}$ state, with ``exotic"
quantum numbers $J^{PC} = 1^{-+}$, can be formed. This state can be
identified as the observed $\pi_1(1600)$ resonance. We suggest that
this is the origin of the present $\pi_1(1600)$ resonance and
propose to look at the $\pi f_1(1285)$ mode in future experiments to
clarify the issue.


\end{abstract}

\pacs{13.75.Lb, 14.20.Dh 11.30.Hv}

\maketitle

\section{Introduction} \label{sec:introduction}

The mesons are described as bound states of quarks and antiquarks in
the classical quark model. Until now, most of the known mesons can
be described very well within the quark model~\cite{Agashe:2014kda}.
However, there is a growing set of experimental observations of
resonance-like structures with quantum numbers which are forbidden
for the quark-antiquark ($q\bar{q}$) system or situated at masses
which cannot be explained by the classical quark
model~\cite{Klempt:2007cp,Brambilla:2014jmp}. From the experimental
side, new observations in the heavy quark sector have reported of
several mesons with nonconventional
features~\cite{Choi:2003ue,Acosta:2003zx,Abazov:2004kp,Ablikim:2013mio,Liu:2013dau,Adolph:2015pws,D0:2016mwd}.

A state with quantum numbers $J^{PC} = 1^{-+}$ cannot be described
as simple quark antiquark pairs~\cite{Amsler:2004ps}. For $J^{PC} =
1^{-+}$ the angular momentum $l$ between the quark and the antiquark
must be even, since $P = -(-1)^l$. The positive $C$-parity then
requires the total quark spin $s$ to be zero, since $C =
(-1)^{l+s}$. This then implies $J = l$ and therefore excludes $J =
1$. But, the quantum numbers of these exotic states could be
obtained within the hybrid configurations by adding a gluonic
excitation to the $q\bar{q}$ pair and such exotic hybrid
configurations should be observed as additional states in the meson
spectrum. In the light quark sector there are three quite
well-established exotic candidates with $J^{PC} = 1^{-+}$:
$\pi_1(1400)$, $\pi_1(1600)$, and $\pi_1(2015)$. Over the past two
decades, both experimental and theoretical sides have put forth many
efforts to investigate these exotic mesons~\cite{Meyer:2015eta}. The
$\pi_1(1600)$ state was observed by the E852 Collaboration in the
$\rho \pi$ channel with the reaction $\pi^- p \to \pi^- \pi^+ \pi^-
p$~\cite{Adams:1998ff,Chung:2002pu}, in the $\eta' \pi$ channel with
the reaction $\pi^- p \to \eta' \pi^- p$~\cite{Ivanov:2001rv}, in
the $f_1(1285) \pi$ channel with the reaction $\pi^- p \to \eta
\pi^+ \pi^- \pi^- p$~\cite{Kuhn:2004en}, and in the $b_1 \pi$
channel with the reaction $\pi^- p \to \pi^+ \pi^- \pi^- \pi^0 \pi^0
p$~\cite{Lu:2004yn}. Later, COMPASS Collaboration at CERN showed
further evidence for $\pi_1(1600)$ in the $\rho \pi$
channel~\cite{Alekseev:2009aa} with mass $M_{\pi_1(1600)} = 1660 \pm
10^{+0}_{-64}$ MeV and a width of $\Gamma_{\pi_1(1600)} = 269 \pm
21^{+42}_{-64}$ MeV. However, the CLAS Collaboration at JLab did not
find the evidence of $\pi_1(1600)$ state through the
photo-production process $\gamma p \to \pi^+ \pi^+ \pi^- (n)_{\rm
missing}$~\cite{Mecking:2003zu,Nozar:2008aa}.

Within different theoretical approaches, there are many
investigations of the light $1^{-+}$ hybrid meson properties in
Refs.~\cite{Isgur:1984bm,Close:1994hc,Page:1998gz,Ebert:2009ub,Kim:2008qh,Dudek:2010wm,Meyer:2010ku,Bellantuono:2014lra}.
However, the calculations of the mass of the lightest $1^{-+}$ meson
in those works are different. For example, in
Ref.~\cite{Meyer:2010ku}, it is found that the $\pi_1(1600)$ could
be the lightest exotic quantum number hybrid meson, while the
results in Ref.~\cite{Bellantuono:2014lra} favor $\pi_1(1400)$ as
the lightest hybrid state. Furthermore, the decay properties of the
$1^{-+}$ hybrid state are studied within the framework of the $QCD$
sum rules in Ref.~\cite{Chen:2010ic} and the chiral corrections to
the $\pi_1(1600)$ state are calculated up to one-loop order in
Ref.~\cite{Zhou:2016saz}. There are also other interpretations that
$\pi_1(1600)$ might be a four-quark state~\cite{Chen:2008qw} or a
molecule/four-quark mixing state~\cite{Narison:2009vj}.

On the basis of the experimental and theoretical studies of the
$1^{-+}$ hybrid mesons, the identification of the $\pi_1(1600)$
state is a debated issue, thus it is still worth studying the
$\pi_1(1600)$ state in different ways.

In this article, we investigate the $\pi_1(1600)$ state in
three-body system of $\pi \bar{K} K^*$ but keep the strong
correlations of the $\bar{K} K^*$ system~\footnote{Note that the
$|\bar{K} K^*>$ state has no well-defined $C$- and $G$-parity, but
it is known that the combination $\frac{1}{\sqrt{2}} (|\bar{K}K^*> +
|K\bar{K}^*>)$ is $C$- and $G$-parity eigenstate with $C=+1$ and
$G=+1$ (see more details in Ref.~\cite{Roca:2005nm}), and
$f_1(1285)$ is a bound state of $\frac{1}{\sqrt{2}} (|\bar{K}K^*> +
|K\bar{K}^*>)$. However, as we shall see later, the output of our
calculation with $|\bar{K}K^*>$ is the same as $\frac{1}{\sqrt{2}}
(|\bar{K}K^*> + |K\bar{K}^*>)$ for $f_1(1285)$. Thus, in this work,
we take only $|\bar{K}K^*>$ for $f_1(1285)$.} which generate
$f_1(1285)$ resonance in the isospin $I = 0$
sector~\cite{Roca:2005nm,Lutz:2003fm}. In such a situation the use
of the fixed center approximation (FCA) to the Faddeev equation is
justified~\cite{Gal:2006cw,Barrett:1999cw,Kamalov:2000iy}. The FCA
to the Faddeev equations has been used with success recently in
Ref.~\cite{Xie:2010ig} for the case of $N\bar{K}K$ system, with
results very similar to those obtained in full Faddeev calculations
in Refs.~\cite{MartinezTorres:2008kh,MartinezTorres:2010zv} and in
the variational estimate in Ref.~\cite{Jido:2008kp}. With FCA to the
Faddeev equations, the $\Delta_{5/2^+}(2000)$ puzzle is solved in
the study of the $\pi$-$(\Delta \rho)_{N_{5/2^-}(1675)}$
system~\cite{Xie:2011uw}. In
Refs.~\cite{Bayar:2013bta,Liang:2013yta,Durkaya:2015wra,Bayar:2015oea},
by taking the FCA to Faddeev equations the three-body systems of
$\rho K \bar{K}$, $\eta K \bar{K}$, $\eta' K \bar{K}$, $\rho D
\bar{D}$, and $\rho D^* \bar{D}^*$ were investigated. Besides, the
$\pi(1300)$ resonance was obtained in the study of
three-pseudoscalar $\pi K \bar{K}$ and $\pi \pi \eta$ coupled system
by solving the Faddeev equations within an approach based on unitary
chiral dynamics~\cite{MartinezTorres:2011vh}. For $2^{-+}$
pseudotensor mesons, it was shown that, in Ref.~\cite{Roca:2011br},
the $\pi_2(1670)$, $\eta_2(1645)$ and $K^*_2(1770)$ can be regarded
as molecules made of a pseudoscalar and a tensor meson, where the
latter is itself made of two vector mesons.

In the present work we will use the FCA to Faddeev equations to
investigate the $\pi \bar{K} K^*$ system. When studied in $s$-wave,
provided the strength of the interactions allows for it, the
$\pi$-$(\bar{K} K^*)_{f_1(1285)}$ system could give rise to the
exotic $\pi_1$ states with quantum numbers $I^{G}(J^{PC}) =
1^-(1^{-+})$. In terms of two-body $\pi \bar{K}$ and $\pi K^*$
scattering amplitudes obtained from the chiral unitary
approach~\cite{Roca:2005nm,Geng:2006yb,Guo:2005wp}, we perform an
analysis of the $\pi$-$(\bar{K} K^*)_{f_1(1285)}$ scattering
amplitude, which will allow us to identify dynamically generated
resonances with the exotic states discussed above.

In the next section, we present the FCA formalism and ingredients to
analyze the $\pi$-$(\bar{K} K^*)_{f_1(1285)}$ system. In Sec. III,
our results and discussions are presented. Finally, a short summary
is given in Sec. IV.

\section{Formalism and ingredients} \label{sec:formalism}

\begin{figure*}[htbp]
\begin{center}
\includegraphics[scale=0.9]{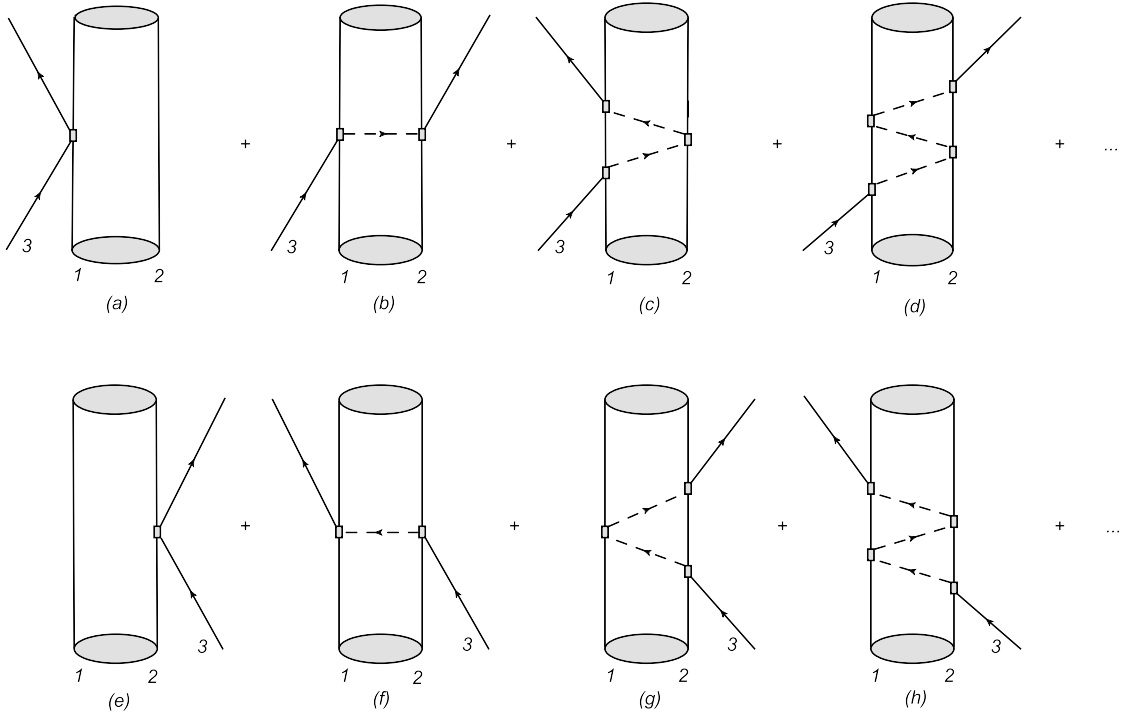}
\caption{Diagrammatic representation of the FCA to Faddeev
equations.} \label{fig:FCA-diagram}
\end{center}
\end{figure*}

The FCA approximation to Faddeev equations assumes a pair of
particles (1 and 2) forming a cluster. Then particle 3 interacts
with the components of the cluster, undergoing all possible multiple
scattering with those components. This is depicted in
Fig.~\ref{fig:FCA-diagram}. In terms of the two partition functions
$T_1$ and $T_2$, which sum all diagrams of the series of
Fig.~\ref{fig:FCA-diagram} that begin with the interaction of
particle 3 with the particle 1 of the cluster ($T_1$), or with the
particle 2 ($T_2$), the FCA equations are
\begin{eqnarray}
T_1 &=& t_1+t_1G_0T_2,  \label{eq:FCA-T1} \\
T_2 &=& t_2+t_2G_0T_1, \label{eq:FCA-T2} \\
T  &=& T_1+T_2, \label{eq:FCA-Ttotal}
\end{eqnarray}
where $T$ is the total scattering amplitude. The amplitudes $t_1$
and $t_2$ represent the unitary scattering amplitudes with coupled
channels for the interactions of particle 3 with particle 1 and 2,
respectively. In the present work, we consider $\bar{K}K^*$ as a
bound state of the $f_1(1285)$, thus $\bar{K}$ and $K^*$ are
particles 1 and 2, respectively. The $\pi$ meson is particle 3.
Then, $t_1$ is the combination of the $I = 1/2$ and $3/2$ unitarized
two-body $\pi K$ scattering amplitude, while $t_2$ is the $I = 1/2$
and $3/2$ unitarized two-body $\pi K^*$ scattering amplitude. In the
above equations, $G_0$ is the loop function for the $\pi$ meson
propagating inside the $(\bar{K} K^*)_{f_1(1285)}$ cluster which is
discussed below. The analysis of the
$\pi$-$(\bar{K}K^*)_{f_1(1285)}$ scattering amplitude will allow us
to study dynamically generated resonances.

For the evaluation of the two body amplitudes $t_1$ and $t_2$ in
terms of the unitary amplitudes in the isospin basis, we need first
to consider the interaction of a $\pi$ and a $\bar{K} K^*$ cluster.
The $\bar{K} K^*$ in isospin zero is written as,
\begin{eqnarray}
|\bar{K}K^*>_{I = 0} =
\frac{1}{\sqrt{2}}|(\frac{1}{2},-\frac{1}{2})> -
\frac{1}{\sqrt{2}}|(-\frac{1}{2},\frac{1}{2})>,
\end{eqnarray}
where the kets in the right-hand side indicate the $I_z$ components
of the particles $\bar{K}$ and $K^*$, $|(I^{\bar{K}}_z ,
I^{K^*}_z)>$.

Following the procedures of Refs.~\cite{Xie:2010ig,Xie:2011uw},
$t_1$ and $t_2$ can be easily obtained in terms of two-body
amplitudes $t_{31}$ and $t_{32}$. Here we write explicitly the case
of $I_{\pi \bar{K} K^*} = I^z_{\pi \bar{K} K^*} = 1$,
\begin{widetext}
\begin{eqnarray}
&& < \pi \bar{K} K^* |t| \pi \bar{K} K^* > = \left(  <11| \bigotimes
\frac{1}{\sqrt{2}}(<(\frac{1}{2},-\frac{1}{2})| -
<(-\frac{1}{2},\frac{1}{2})|) \right) (t_{31} + t_{32}) \left( |11>
\bigotimes  \frac{1}{\sqrt{2}}(|(\frac{1}{2},-\frac{1}{2})> -
|(-\frac{1}{2},\frac{1}{2})>)  \right) \nonumber \\
&=& \left( \frac{1}{\sqrt{2}}
<(\frac{3}{2}\frac{3}{2},-\frac{1}{2})| -  \frac{1}{\sqrt{6}}
<(\frac{3}{2}\frac{1}{2},\frac{1}{2})| - \frac{1}{\sqrt{3}}
<(\frac{1}{2}\frac{1}{2},\frac{1}{2})| \right) t_{31} \left(
\frac{1}{\sqrt{2}} |(\frac{3}{2}\frac{3}{2},-\frac{1}{2})> -
\frac{1}{\sqrt{6}} |(\frac{3}{2}\frac{1}{2},\frac{1}{2})> -
\frac{1}{\sqrt{3}} |(\frac{1}{2}\frac{1}{2},\frac{1}{2})> \right) +
\nonumber \\
&& \left( \frac{1}{\sqrt{6}} <(\frac{3}{2}\frac{1}{2},\frac{1}{2})|
+ \frac{1}{\sqrt{3}} <(\frac{1}{2}\frac{1}{2},\frac{1}{2})| -
\frac{1}{\sqrt{2}} <(\frac{3}{2}\frac{3}{2},-\frac{1}{2})| \right)
t_{32} \left( \frac{1}{\sqrt{6}}
|(\frac{3}{2}\frac{1}{2},\frac{1}{2})> + \frac{1}{\sqrt{3}}
|(\frac{1}{2}\frac{1}{2},\frac{1}{2})> - \frac{1}{\sqrt{2}}
|(\frac{3}{2}\frac{3}{2},-\frac{1}{2})> \right),
\end{eqnarray}
\end{widetext}
where the notation followed in the last term for the states is
$|(I_{\bar{K} \pi} I^z_{\bar{K}\pi}, I^z_{K^*})>$ for $t_{31}$,
while $|(I_{K^* \pi}I^z_{K^* \pi}, I^z_{\bar{K}})>$ for $t_{32}$.
This leads to the following amplitudes~\footnote{Because of charge
conjugation symmetry, the amplitude for $\pi \bar{K}$ scattering is
the same as that for $\pi K$ scattering.} for the single-scattering
contribution [Figs.~\ref{fig:FCA-diagram} (a) and (e)],
\begin{eqnarray}
t_1
&=&\frac{2}{3}t_{\pi K}^{I =3/2} +\frac{1}{3}t_{\pi K}^{I=1/2}, \label{eq:t1}\\
t_2 &=& \frac{2}{3}t_{\pi K^*}^{I = 3/2} + \frac{1}{3}t_{\pi K^*}^{I
= 1/2}. \label{eq:t2}
\end{eqnarray}

On the other hand, it is worth noting that the argument of the total
scattering amplitude $T$ is the total invariant mass $s$ of the
three-body system, while the arguments of $t_1$ and $t_2$ are $s_1$
and $s_2$, where $s_i$ $(i=1,2)$ is the invariant mass of the
interaction particle $\pi$ and the particle $\bar{K}$ $(i = 1)$ or
$K^*$ $(i = 2)$. The value of $s_i$ is given by
\begin{eqnarray}
s_1 &=& m^2_\pi + m^2_{\bar{K}} + \frac{M_R^2 + m_{\bar{K}}^2 -
m_{K^*}^2}{2M_R^2}(s - m^2_{\pi} -
M_R^2), \label{eq:s1} \\
s_2 &=& m^2_\pi + m^2_{K^*} + \frac{M_R^2 + m_{K^*}^2 -
m_{\bar{K}}^2}{2M_R^2}(s - m^2_{\pi} - M_R^2), \label{eq:s2}
\end{eqnarray}
where $M_R$ is the mass of the $f_1(1285)$ state, and we take $M_R$
= 1281.3 MeV.

Then, following the approach developed in
Refs.~\cite{Roca:2010tf,YamagataSekihara:2010qk}, we can easily
obtain the $S$-matrix for the single-scattering term
[Fig.~\ref{fig:FCA-diagram} (a) and (e)] as
\begin{eqnarray}
S^{(1)} &= & S^{(1)}_1 + S^{(1)}_2 \nonumber \\
&=& \frac{(2\pi)^4}{V^2} \delta^4(k+k_R-k'-k_R')
\frac{1}{\sqrt{2\omega_\pi}}
\frac{1}{\sqrt{2\omega'_\pi}} \nonumber \\
&& \!\!\! \times  \left ( -it_1 F_R \Big[
\frac{m_{K^*}(\vec{k}-\vec{k}')}{m_{\bar{K}} + m_{K^*}} \Big]
\frac{1}{\sqrt{2\omega_{\bar{K}}}}
\frac{1}{\sqrt{2\omega'_{\bar{K}}}} \right. \nonumber \\
&& \left. - it_2 F_R \Big[
\frac{m_{\bar{K}}(\vec{k}-\vec{k}')}{m_{\bar{K}} + m_{K^*}} \Big]
\frac{1}{\sqrt{2\omega_{K^*}}} \frac{1}{\sqrt{2\omega'_{K^*}}}
\right ), \label{eq:single-amplitude}
\end{eqnarray}
where $V$ stands for the volume of a box in which the states are
normalized to unity, while $k$, $k'$ ($k_R$, $k_R'$) refer to the
momentum of the initial, final scattering particle ($R$ for the
cluster), $\omega_\pi$ ($\omega_{\bar{K}}$, $\omega_{K^*}$)  and
$\omega'_\pi$ ($\omega'_{\bar{K}}$, $\omega'_{K^*}$) are the
energies of the initial and final scattering particles.

In Eq.~\eqref{eq:single-amplitude}, $F_R$ is the form factor of
$f_1(1285)$ as a bound state of $\bar{K} K^*$. This form factor was
taken to be unity neglecting the $\vec{k}$, $\vec{k}'$ momentum in
Refs.~\cite{Roca:2010tf,YamagataSekihara:2010qk} where only states
below threshold were considered. To consider states above threshold,
we project the form factor into the $s$-wave, the only one that we
consider. Hence
\begin{eqnarray}\label{eq:ffs}
F_R \Big[
\frac{m_{K^*}(\vec{k}-\vec{k}')}{m_{\bar{K}} + m_{K^*}} \Big] & \Rightarrow  FFS_1(s) =
\frac{1}{2} \int_{-1}^{1}F_R(k_1)d({\rm cos}\theta), \\
F_R \Big[
\frac{m_{\bar{K}}(\vec{k}-\vec{k}')}{m_{\bar{K}} + m_{K^*}} \Big] & \Rightarrow  FFS_2(s) =
\frac{1}{2} \int_{-1}^{1}F_R(k_2)d({\rm cos}\theta),
\end{eqnarray}
with
\begin{eqnarray}
k_1 &=& \frac{m_{K^*}}{m_{\bar{K}} + m_{K^*}} k \sqrt{2(1 - {\rm
cos}\theta)}, \\
k_2 &=& \frac{m_{\bar{K}}}{m_{\bar{K}} + m_{K^*}} k \sqrt{2(1 - {\rm
cos}\theta)},
\end{eqnarray}
and
\begin{align}
k = \frac{\sqrt{(s - (m_{\bar{K}} + m_{K^*} + m_\pi)^2)(s -
(m_{\bar{K}} + m_{K^*} - m_\pi)^2)}}{2\sqrt{s}},
\end{align}
is the module of the momentum of the $\pi$ meson in $\pi \bar{K}
K^*$ center-of-mass frame when $\sqrt{s}$ is above the threshold of
the $\pi \bar{K} K^*$ system; otherwise, $k$ equals zero. The
expression of $F_R$ is given below.

The double scattering contributions are from
Figs.~\ref{fig:FCA-diagram} (b) and (f). The expression for the
$S$-matrix for the double scattering [$S^{(2)}_2 = S^{(2)}_1$] is
given by
\begin{eqnarray}\label{eq:double-amplitude}
S^{(2)}_1 &= & -it_1 t_2 \frac{(2\pi)^4}{V^2} \delta^4(k+k_R-k'-k_R') \nonumber \\
&& \times \frac{1}{\sqrt{2\omega_\pi}} \frac{1}{\sqrt{2\omega'_\pi}} \frac{1}{\sqrt{2\omega_{\bar{K}}}} \frac{1}{\sqrt{2\omega'_{\bar{K}}}} \frac{1}{\sqrt{2\omega_{K^*}}}
\frac{1}{\sqrt{2\omega'_{K^*}}} \nonumber \\
&& \times \int \frac{d^3q}{(2\pi)^3} F_R(q)
\frac{1}{{q^0}^2-{\vec{q}}^2-m^2_{\pi} + i \epsilon},
\end{eqnarray}
with
\begin{eqnarray}
q^0 = \frac{s+m^2_\pi -M^2_R}{2\sqrt{s}}.
\end{eqnarray}

One of the ingredients in the calculation is the form factor
$F_R(q)$ for the bound state $f_1(1285)$ of a pair of $\bar{K} K^*$.
Following the approach of
Refs.~\cite{Roca:2010tf,YamagataSekihara:2010qk}, we can easily get
the following expression for the form factor $F_R(q)$,
\begin{eqnarray}\label{eq:Fq}
F_R(q) &=& \frac{1}{N} \int_{|\vec{p}| < \Lambda, ~ |\vec{p} - \vec{q}| < \Lambda} d^3 \vec{p} \frac{1}{2\omega_{\bar{K}}(\vec{p})} \frac{1}{2\omega_{K^*}(\vec{p})} \nonumber \\
&& \times \frac{1}{M_R - \omega_{\bar{K}}(\vec{p}) - \omega_{K^*}(\vec{p})} \frac{1}{2\omega_{\bar{K}}(\vec{p} - \vec{q})}\frac{1}{2\omega_{K^*}(\vec{p}-\vec{q})} \nonumber \\
&& \times \frac{1}{M_R - \omega_{\bar{K}}(\vec{p} - \vec{q}) - \omega_{K^*}(\vec{p}-\vec{q})},
\end{eqnarray}
where the normalization factor $N$ is
\begin{eqnarray}\label{eq:normalizationfq}
N \! \! = \!\! \int_{|\vec{p}| < \Lambda} d^3 \vec{p} \Big( \frac{1}{2\omega_{\bar{K}}(\vec{p})} \frac{1}{2\omega_{K^*}(\vec{p})}
\frac{1}{M_R - \omega_{\bar{K}}(\vec{p}) - \omega_{K^*}(\vec{p})}\Big)^2.
\end{eqnarray}
The parameter $\Lambda$ is used to regularize the loop functions in
the chiral unitary approach~\cite{Roca:2005nm}.

In this work we take $\Lambda$ around $990$ MeV such that the
$f_1(1285)$ is obtained~\cite{Roca:2005nm}. The condition
$|\vec{p}-\vec{q}| < \Lambda$ implies that the form factor is
exactly zero for $q > 2\Lambda$. Therefore the integration in
Eq.~(\ref{eq:Fq}) has upper limit of $2\Lambda$.

\begin{figure}[htbp]
\centering
\includegraphics[scale=0.9]{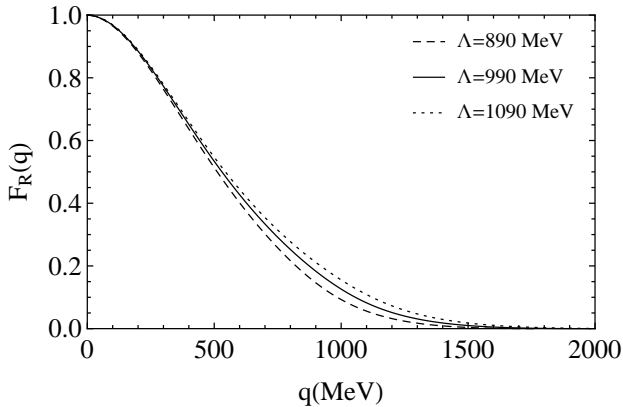}
\caption{Form factor of the $f_1(1285)$ as a $\bar{K} K^*$ bound state.} \label{fig:Fq}
\end{figure}

We show the form factor $F_R(q)$ in Fig.~\ref{fig:Fq} with $\Lambda
= 890$, $990$, and $1090$ MeV. From Fig.~\ref{fig:Fq} we see that
the form factor $F_R(q)$ is not sensitive to the value of $\Lambda$,
especially for $q < 600$ MeV, and we find that the results of the
total scattering amplitude $T$ are very similar with $\Lambda = 990
\pm 100$ MeV, hence we take $\Lambda = 990$ MeV in the following
such that the $f_1(1285)$ is obtained~\cite{Roca:2005nm}.

\begin{figure}[htbp]
\centering
\includegraphics[scale=0.6]{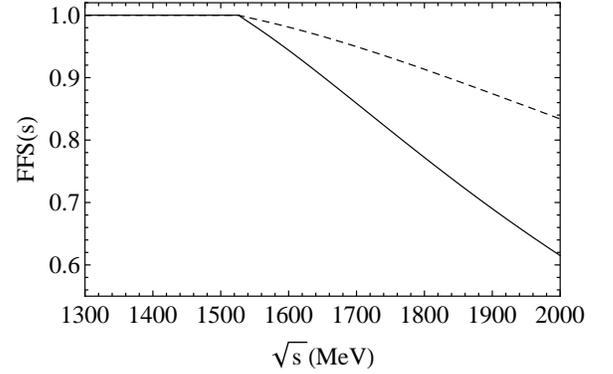}
\caption{Form factor for the single-scattering contribution.} \label{fig:FFS}
\end{figure}

With the results of $F_R(q)$, we can easily calculate the form
factors $FFS_i(s)$ for single scattering. In Fig.~\ref{fig:FFS}, we
show the projection over the $s$-wave of the form factor for the
single scattering contribution as a function of the total invariant
mass of the $\pi \bar{K} K^*$ system. The solid and dashed curves
are the results of $FFS_1$ and $FFS_2$, respectively. We see that
the $FFS_1$ and $FFS_2$ are very close to one below $\sqrt{s} =
1800$ MeV, which indicates that the corrections from these two form
factors are very small and only affect moderately the results of $T$
beyond $1800$ MeV.

Before proceeding further, we examine the normalization for the $S$
matrix, which is given by
\begin{eqnarray}\label{eq:stotal}
S &=& -iT \frac{(2\pi)^4}{V^2} \delta^4(k+k_R-k'-k_R') \nonumber \\
 && \times \frac{1}{\sqrt{2\omega_\pi}} \frac{1}{\sqrt{2\omega'_\pi}} \frac{1}{\sqrt{2\omega_{f_1(1285)}}} \frac{1}{\sqrt{2\omega'_{f_1(1285)}}}.
\end{eqnarray}

By comparing Eq.~\eqref{eq:stotal} with
Eq.~\eqref{eq:single-amplitude} for the single scattering and
Eq.~\eqref{eq:double-amplitude} for the double scattering, we see
that we have to give a weight to $t_1$ and $t_2$ such that
Eqs.~\eqref{eq:single-amplitude} and \eqref{eq:double-amplitude} get
the weight factors that appear in the general formula of
Eq.~\eqref{eq:stotal}. This is achieved by replacing
\begin{eqnarray}\label{eq:t1t2normalization}
t_1 \to \tilde{t}_1  & = & t_1 \sqrt{\frac{2\omega_{f_1(1285)}}{2\omega_{\bar{K}}}} \sqrt{\frac{2\omega'_{f_1(1285)}}{2\omega'_{\bar{K}}}},   \\
t_2 \to \tilde{t}_2  & = & t_2 \sqrt{\frac{2\omega_{f_1(1285)}}{2\omega_{K^*}}} \sqrt{\frac{2\omega'_{f_1(1285)}}{2\omega'_{K^*}}}.
\end{eqnarray}

By solving Eqs.~\eqref{eq:FCA-T1} and \eqref{eq:FCA-T2} and summing the two partitions $T_1$ and $T_2$, we get
\begin{eqnarray}
T  =  \frac{\tilde{t}_1  +  \tilde{t}_2  +  2\tilde{t}_1\tilde{t}_2 G_0}{1-\tilde{t}_1\tilde{t}_2 G_0^2}  +  \tilde{t}_1[FFS_1  - 1]  +  \tilde{t}_2[FFS_2  -  1], \label{eq:tpif1}
\end{eqnarray}
where $G_0$ depends on the invariant mass square $s$ and is given by
\begin{eqnarray}
G_0(s) = \frac{1}{2\omega_{f_1(1285)}} \int \!\! \frac{d^3\vec{q}}{(2\pi)^3} F_R(q) \frac{1}{{q^0}^2 - {\vec{q}}^2 - m^2_\pi + i \epsilon}.\label{gzero}
\end{eqnarray}

\begin{figure}[htbp]
\centering
\includegraphics[scale=0.6]{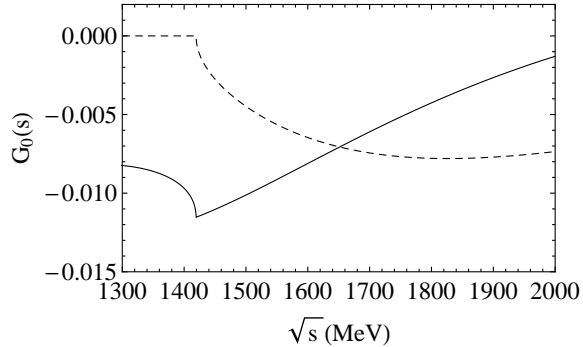}
\caption{Real (solid line) and imaginary (dashed line) parts of the $G_0$ function.} \label{fig:gzero}
\end{figure}

In Fig.~\ref{fig:gzero}, we show the real and imaginary parts of the
$G_0$ as a function of the invariant mass of the $\pi \bar{K} K^*$
system.

\section{RESULTS AND DISCUSSION}

To perform the evaluation of Faddeev equations under the FCA, we
need the calculation of the two-body interaction amplitudes ($t_1$
and $t_2$) of $\pi \bar{K}$ and $\pi K^*$, which are investigated in
Refs.~\cite{Roca:2005nm,Geng:2006yb,Guo:2005wp} as mentioned before.
These two-body scattering amplitudes depend on the subtraction
constants $a_{\pi \bar{K}}$ and $a_{\pi K^*}$, which are assumed as
effective parameters in our calculation. We take them as used in
Refs.~\cite{Geng:2006yb,Guo:2005wp}: $a_{\pi K^*} = -1.85$ and $\mu
= 1000$ MeV for $I_{\pi K^*} = 1/2$; $a_{\pi K} = -1.38$ and $\mu =
m_K$ for $I_{\pi K} = 1/2$; $a_{\pi K} = -4.64$ and $\mu = m_K$ for
$I_{\pi K} = 3/2$. Then we calculate the total scattering amplitude
$T$ and associate the peaks/bumps in the modulus squared $|T|^2$ to
resonances.

In Ref.~\cite{Geng:2006yb}, only the $\pi K^*$ interaction in
$I_{\pi K^*} = 1/2$ sector was studied where two $K_1(1270)$ states
were obtained. In this work we need also the parameter $a_{\pi K^*}$
for the case of $I_{\pi K^*} = 3/2$, which is taken the same as for
$I_{\pi K^*}=1/2$ as used in Ref.~\cite{Geng:2006yb}.

In the FCA, we keep the wave function of the cluster unchanged by
the presence of the third particle. In order to estimate
uncertainties of the FCA due to this ¡°frozen¡± condition we admit
that the wave function of the cluster could be modified by the
presence of the third particle, which is the normal situation in a
full Faddeev calculation. Indeed, $\pi f_1(1285)$ may couple to
other $s$-wave meson-meson channels, such as $\pi$ meson and other
excited $f_1$ states or $\bar{K}$ meson and $K_1$ states. However,
other excited $f_1$ states may not have large $\bar{K} K^*$
component~\footnote{One might think that the inclusion of
$h_1(1380)$ and $b_1(1235)$ states might improve the situation,
since those resonances couple also dominantly to the $\bar{K}K^*$
channel~\cite{Lutz:2003fm}. However, the quantum numbers of
$h_1(1380)$ and $b_1(1235)$ are different with $f_1(1285)$. The
transition between $\pi$-$(\bar{K}K^*)_{h_1(1380)}$,
$\pi$-$(\bar{K}K^*)_{b_1(1235)}$ and
$\pi$-$(\bar{K}K^*)_{f_1(1285)}$ should be zero.} or the thresholds
of these channels are far from the energy region we considered.
Furthermore, including such contributions, the
$\pi$-$(\bar{K}K^*)_{f_1(1285)}$ scattering amplitude would become
more complex due to additional parameters from the non-diagonal
transitions, and we cannot determine or constrain these parameters.
Hence, we will leave these contributions to future studies when more
experimental data become available. For the sake of simplicity we do
not include other channels in our calculation.

\begin{figure}[htbp]
\centering
\includegraphics[scale=0.9]{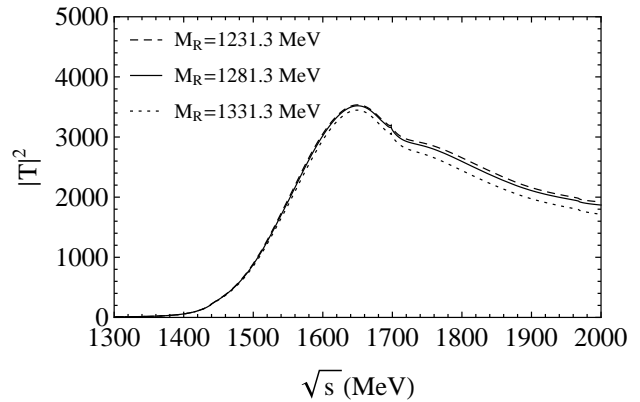}
\caption{Modulus squared of the $\pi \bar{K} K^*$ three-body
scattering amplitude.} \label{Fig:tsquare}
\end{figure}

As pointed before, the form factor, $F_R(q)$, is not sensitive to
the value of $\Lambda$. Then, in order to quantify uncertainties of
the FCA, we perform calculations with different values of $M_R$. In
Fig.~\ref{Fig:tsquare}, we show the modulus squared of the total
$\pi$-$(\bar{K} K^*)_{f_1(1285)}$ scattering amplitude with $M_R =
1231.3$, $1281.3$, and $1331.3$ MeV, where we see a clear bump
structure around $\sqrt{s} \sim 1650$ MeV for the three cases. From
the PDG~\cite{Agashe:2014kda}, this structure can be assigned to
$\pi_1(1600)$, with mass $1660$ MeV. Furthermore, taking $\sqrt{s} =
1660$ MeV we get $\sqrt{s_1} = 792$ MeV and $\sqrt{s_2} = 1244$ MeV
from Eqs.~\eqref{eq:s1} and \eqref{eq:s2}. At these energy points,
the interactions of $\pi \bar{K}$ and $\pi K^*$ are strong enough to
produce the $\pi_1(1600)$ state.

Note that the location of the peak is quite stable against variation
of the parameters of $a_{\pi \bar{K}}$ and $a_{\pi K^*}$ in the
ranges of values to reproduce the results of
Refs.~\cite{Geng:2006yb,Guo:2005wp} within uncertainties. This may
indicate that the $\pi_1(1600)$ state can be generated from $\pi
f_1(1285)$ where $f_1(1285)$ is present in the $\bar{K}K^*$
interaction. This may be the origin of the $\pi_1(1600)$ state and
the future measurements about the $\pi f_1(1285)$ mode can be used
to test our finding here.

On the other hand, from Fig.~\ref{Fig:tsquare} we see that there is
no any bump structure around $\sqrt{s} \sim 1400$ MeV, which can be
assigned as the $\pi_1(1400)$ state. This may indicate that the
$\pi_1(1400)$ can not be dynamically generated from the $\pi
f_1(1285)$ interaction.

\section{Summary}

In this work, we have performed a Faddeev calculation for the
$\pi$-$f_1(1285)$ system treating $f_1(1285)$ state as a $\bar{K}
K^*$ bound state as found in previous studies of the $\bar{K}$-$K^*$
system~\cite{Lutz:2003fm,Roca:2005nm}. We have used the FCA to
describe the $\pi$-$(\bar{K} K^*)_{f_1(1285)}$ system in terms of
the two-body interactions, $\pi \bar{K}$ and $\pi K^*$, provided by
the chiral unitary approach as investigated in
Refs.~\cite{Geng:2006yb,Guo:2005wp}. There is a clear and stable
bump structure around $1650$ MeV in the module squared of the total
scattering amplitude indicating the formation of a resonant $\pi
\bar{K} K^*$ state around this energy. This state has ``exotic"
quantum numbers $J^{PC} = 1^{-+}$. From PDG, we can associated this
resonance to the exotic $\pi_1(1600)$ state with mass $1660$ MeV and
large uncertainties for the width~\cite{Agashe:2014kda}. This may be
the origin of the $\pi_1(1600)$ resonance  that is treated as a
hybrid state in Refs.~\cite{Chen:2010ic,Zhou:2016saz}, a four-quark
state in Ref.~\cite{Chen:2008qw} or a molecule/four-quark mixing
state in Ref.~\cite{Narison:2009vj}. Future measurements about the
$\pi f_1(1285)$ mode can be used to test our calculations and
clarify the issue.

\section*{Acknowledgments}

This work is partly supported by the National Basic Research Program (973 Program Grant No. 2014CB845406), by the National Natural Science
Foundation of China under Grant No. 11475227 and the Youth Innovation Promotion Association CAS (No. 2016367).

\clearpage

\end{document}